\documentclass[a4paper,twocolumn,aps,floatfix,citeautoscript,superscriptaddress,nobibnotes]{revtex4-1} 

\usepackage{graphicx}
\usepackage{textcomp}
\usepackage{amsfonts,amsmath,latexsym,amssymb,gensymb}
\usepackage{bm}
\usepackage{color}

\begin{document} 
	
\newcommand{\Dy}{Dy$_2$Ti$_2$O$_7$}
\newcommand{\Ho}{Ho$_2$Ti$_2$O$_7$}
\newcommand{\Tb}{Tb$_2$Ti$_2$O$_7$}
\newcommand{\Ir}{Pr$_2$Ir$_2$O$_7$}
\newcommand{\R}{R$_2$M$_2$O$_7$}
\newcommand{\HoIr}{Ho$_2$Ir$_2$O$_7$}
\newcommand{\OO}{O$^{-2}$}

\newcommand{\tetrahedron}{
  \mathchoice
    {\includegraphics[height=1.4ex]{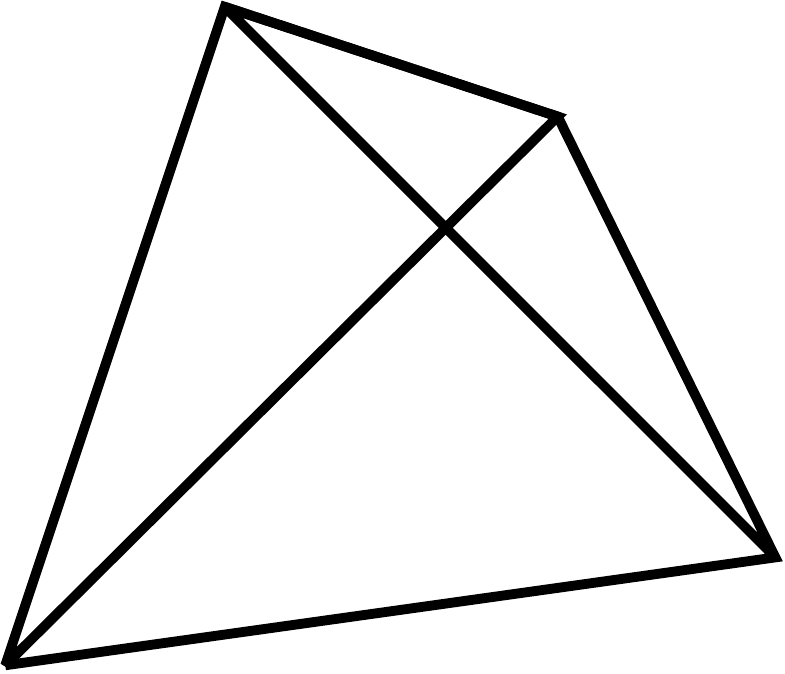}} 
    {\includegraphics[height=1.4ex]{fig0.pdf}} 
    {\includegraphics[height=1.2ex]{fig0.pdf}} 
    {\includegraphics[height=.9ex]{fig0.pdf}} 
}

\makeatletter
\renewcommand{\fnum@figure}{\textbf{Figure~\thefigure}}
\makeatother

\title{Monopole matter from magnetoelastic coupling in the Ising pyrochlore}

\author{D. Slobinsky}
\email[]{These authors contributed equally.}
\affiliation{{Instituto de F\'{\i}sica de L\'{\i}quidos y Sistemas Biol\'ogicos (IFLYSIB), UNLP-CONICET, La Plata, Argentina.}}
\affiliation{Departamento de Ingenier\'ia Mec\'anica, Facultad Regional La Plata, Universidad Tecnol\'ogica Nacional, Av. 60 Esq. 124, 1900 La Plata, Argentina}

\author{L. Pili}
\email[]{These authors contributed equally.}
\affiliation{{Instituto de F\'{\i}sica de L\'{\i}quidos y Sistemas Biol\'ogicos (IFLYSIB), UNLP-CONICET, La Plata, Argentina.}}
\affiliation{Departamento de F\'\i{}sica, Facultad de Ciencias Exactas, Universidad Nacional de La Plata, 1900 La Plata, Argentina.}

\author{G. Baglietto}
\email[]{These authors contributed equally.}
\affiliation{{Instituto de F\'{\i}sica de L\'{\i}quidos y Sistemas Biol\'ogicos (IFLYSIB), UNLP-CONICET, La Plata, Argentina.}}

\author{S. A. Grigera}
\affiliation{{Instituto de F\'{\i}sica de L\'{\i}quidos y Sistemas Biol\'ogicos (IFLYSIB), UNLP-CONICET, La Plata, Argentina.}}
\affiliation{Departamento de F\'\i{}sica, Facultad de Ciencias Exactas, Universidad Nacional de La Plata, 1900 La Plata, Argentina.}
\affiliation{School of Physics and Astronomy, University of St. Andrews, St. Andrews KY16 9SS, United Kingdom}

\author{R. A. Borzi}
\email[Corresponding author e-mail: ]{borzi@fisica.unlp.edu.ar}
\affiliation{{Instituto de F\'{\i}sica de L\'{\i}quidos y Sistemas Biol\'ogicos (IFLYSIB), UNLP-CONICET, La Plata, Argentina.}}
\affiliation{Departamento de F\'\i{}sica, Facultad de Ciencias Exactas, Universidad Nacional de La Plata, 1900 La Plata, Argentina.}

\date{\today}

 \begin{abstract}
Ising models on a pyrochlore oxide lattice are usually associated with spin ice materials and “magnetic monopoles”. Ever more often effects connecting magnetic and elastic degrees of freedom are reported on these and other related frustrated materials. Here we extend a spin-ice Hamiltonian to include coupling between spins and the \OO~ions mediating superexchange; we call it the Magnetoelastic Spin Ice model (MeSI).  There has been a long search for a model in which monopoles would spontaneously become the building blocks of new ground-states: the MeSI Hamiltonian is such a model. In spite of its simplicity and classical approach, it describes (both spin and oxygen lattice) the double-layered monopole crystal observed in \Tb. Remarkably, the dipolar electric moment of single monopoles emerges as a probe for magnetism. As an example we show that, in principle, pinch points related to Coulomb phases could be detected in association with the \OO-ion displacements.
\end{abstract}

\maketitle

\section{Introduction}

It is remarkable that even one of the simplest interacting models in condensed matter physics (the Ising model) can lead to some of the phenomena in geometrically frustrated magnetism that has kept busy part of our community for the past decades. At the core 
of the new emergent physics --massively degenerate ground states with  critical-like spin correlations, exotic excitations, artificial electrostatics, and very peculiar dynamics~\cite{balents2010,moessner1998,mostame2014tunable,Henley2010,lacroix2011,diep2004,moessner03}-- is the strategic choice of the spin lattice structure 
so that pairwise interactions compete rather than collaborate.

The pyrochlore structure (Fig.~\ref{figTetra}a)) is a prominent example among these ``frustrated" lattices, with spin ice canonical materials \Dy~and \Ho~ 
as some of its most notable examples ~\cite{diep2004,lacroix2011,moessner03,castelnovo2012annual,bramwell2001sci,bramwell2020history}. Their effective residual magnetic entropy is similar to that of water ice, and the source of their collective name. The configurations of Ising spins in the lowest energy states of these materials
~\cite{isakov2005why,Castelnovo2008,melko2004} can be described by a lattice gauge field that fluctuates like an electric field in vacuum~\cite{bram2001pinch,isakov2004bdipcorr,fennell2007pinchpKagome,Fennell2009coulombphase,Henley2010}.  
The combination of this ``Coulomb Phase" with non-negligible dipolar interactions leads in turn to the emergence of local magnetic excitations: the ``monopoles". They sit in the centres of the tetrahedra that make the pyrochlore lattice, and interact through Coulomb forces like electrical charges~\cite{Castelnovo2008,morris2009}.  As illustrated in Fig.~\ref{figTetra}a) there are different types of these magnetic charge-like quasiparticles (eight ``single" monopoles, two ``double" ones). Monopoles are responsible for the very peculiar dynamics of spin ices at low temperatures~\cite{snyder2001spin,Jaubert2009nat,Slobinsky2010,paulsen2014quenches}. Also, under different conditions, they can act as building blocks for different ``monopole phases"~\cite{Castelnovo2008,castelnovo2012annual,Borzi2013,Guruciaga2014,Sazonov2012_double} that have been studied theoretically or experimentally. 
In general, dense ``monopole matter" was forced to appear by resourcing to somewhat artificial conditions ~\cite{Borzi2013,Bbartlett2014,xie2015,Jaubert2015spin,Slobinsky2018}, freezing spin fluctuations~\cite{Castelnovo2008,Guruciaga2014}, imposing out of equilibrium situations~\cite{Udagawa2016}, or breaking some symmetry of the system~\cite{Lefrancois2017,Jaubert2015spin,Raban2019multiple,Guruciaga2016,Guruciaga2019,Slobinsky2019polarized}. 
It can be proved to be impossible to obtain the most general monopole liquid solely from pairwise interactions  \cite{Slobinsky2018}, leaving unanswered a fundamental question that we pursuit to respond here: how can monopole matter be thermodynamically stable in real materials without explicitly breaking any symmetry?

Central to this question and to this work is the interplay between magnetic and elastic degrees of freedom. Since it is the precise geometry of the lattice the one that balances out the pairwise spin interactions, geometrically frustrated systems can be quite susceptible to spontaneous deformation~\cite{tchernyshyov2011lacroix, yamashita2000spin,becca2002peierls,jia2006spin,penc2004,bergman2006models,pili2019two}. Regarding Ising pyrochlores that remain disordered at the lowest temperatures, this coupling is responsible for structural fluctuations~\cite{Ruff2007structuralfluct}, giant magnetostriction~\cite{belov1983giant,Ruff2010magnetoelastics}, and composite magnetoelastic excitations in \Tb~\cite{Fennell2014magnetoelastic}. It seems to be much smaller in the canonical spin ices~\cite{stoter2020extremely,edberg2019dipolar}, but may explain subtle effects shaping the zero magnetic field ($\textbf{h}=0$), and $\textbf{h}||$[111] phase diagrams of \Dy~and \Ho~\cite{Borzi2016}, dynamics~\cite{Hornu2020elecdipol}, and the observed magnetic avalanches~\cite{Slobinsky2010,Fenn2005dynamics,fennell2002field}. Khomskii \cite{Khomskii2012} was the first to notice that, in spin ice, spin configurations related to single monopoles are necessarily accompanied by local distortions that result in an electric dipole. These dipoles can interact with an external electric field~\cite{Khomskii2012} or among themselves~\cite{Jaubert2015prb}, changing the energy balance.

Inverting Khomskii's line of reasoning, we demonstrate in this work that magnetoelasticity can be the keystone for monopole stabilization in pyrochlore oxides. We modify the nearest neighbours spin ice Hamiltonian  in the simplest possible way to include a coupling to the lattice of \OO-ions sitting near the centre of tetrahedra (see Fig.~\ref{figTetra}). In the regime of strong coupling, lattice distortions turn the eight types of single monopoles into stable, atomic-like constituents of novel ground states. 
In our model, distortions are not just dummy variables but dynamic degrees of freedom. We can contrast their behaviour with that of real materials, employ them as probes to investigate the underlying magnetism, or ---in a multiferroic fashion--- to control the material properties using electric fields. 

In spite of its simplicity, and building on the previous works of Jaubert and Moessner~\cite{Jaubert2015prb} and Sazonov and collaborators~\cite{Sazonov2012_double}, our  Magnetoelastic Spin Ice (MeSI) model allows to understand in a new manner the formation of a double-layered monopole crystal in \Tb~ with field applied along [110], and to contrast the \OO-distortions with those suggested previously~\cite{Sazonov2012_double}.
The model's output is compatible with the power-law spin correlations observed in \Tb\ at zero field~\cite{Fennell2012}, and gives some clues on the half moons measured in neutron diffraction patterns at finite energy~\cite{guitteny2013,Fennell2014magnetoelastic}.

Although we concentrate on the strong coupling limit, we expect the MeSI model to be a convenient tool to study other systems, in particular spin ices. Incorporating the lattice degrees of freedom  may open the way to the survey or design of the electrical properties on Ising pyrochlores, or teach us how to probe other properties through them (as it has been done in some pioneering works in spin-ice~\cite{katsufuji2004,saito2005magnetodielectric,liu2013multiferroicity,grams2014}).

The article is organised as follows. We begin the Results Section by introducing and justifying the extended magnetoelastic model (Subsection~\ref{sec:Mod}). We then show how the MeSI model stabilizes a Monopole Liquid (Subsection~\ref{sec:ML}) . 
This massively degenerate perfect paramagnet will serve as the basis from which the other cases of study will follow through small perturbations. Including attraction between monopoles of equal charge will lead to a phase comparable to the ``jellyfish" or ``spin slush"~\cite{Udagawa2016,Rau2016}, with half moons in the neutron structure factor. Correspondingly, Coulomb-like attraction gives rise to  a Zincblende Monopole Crystal with magnetic moment fragmentation~\cite{Bbartlett2014,Jaubert2015spin} (Subsection~\ref{sec:MonCrys}) . We will see that the deformed O lattice fluctuates with the fragmented magnetic moments; the pinch points in its structure factor would be, in principle, detectable using diffuse x-ray diffraction. 
The explicit breaking of a symmetry by a magnetic field is addressed in Subsection~\ref{sec:DobLay} (the double-layered Monopole Crystal~\cite{Sazonov2012_double,Jaubert2015prb}). The Discussion (Section~\ref{sec:Discussion}) shows some striking results concerning the structure factors of different phases, evaluates the possibility of observing these effects, and suggests possible new avenues of research.

\section{Results}

\subsection{A model for magnetoelasticity in Ising pyrochlores}\label{sec:Mod}

\begin{figure}[htp]
    \includegraphics[width=1.00\columnwidth]{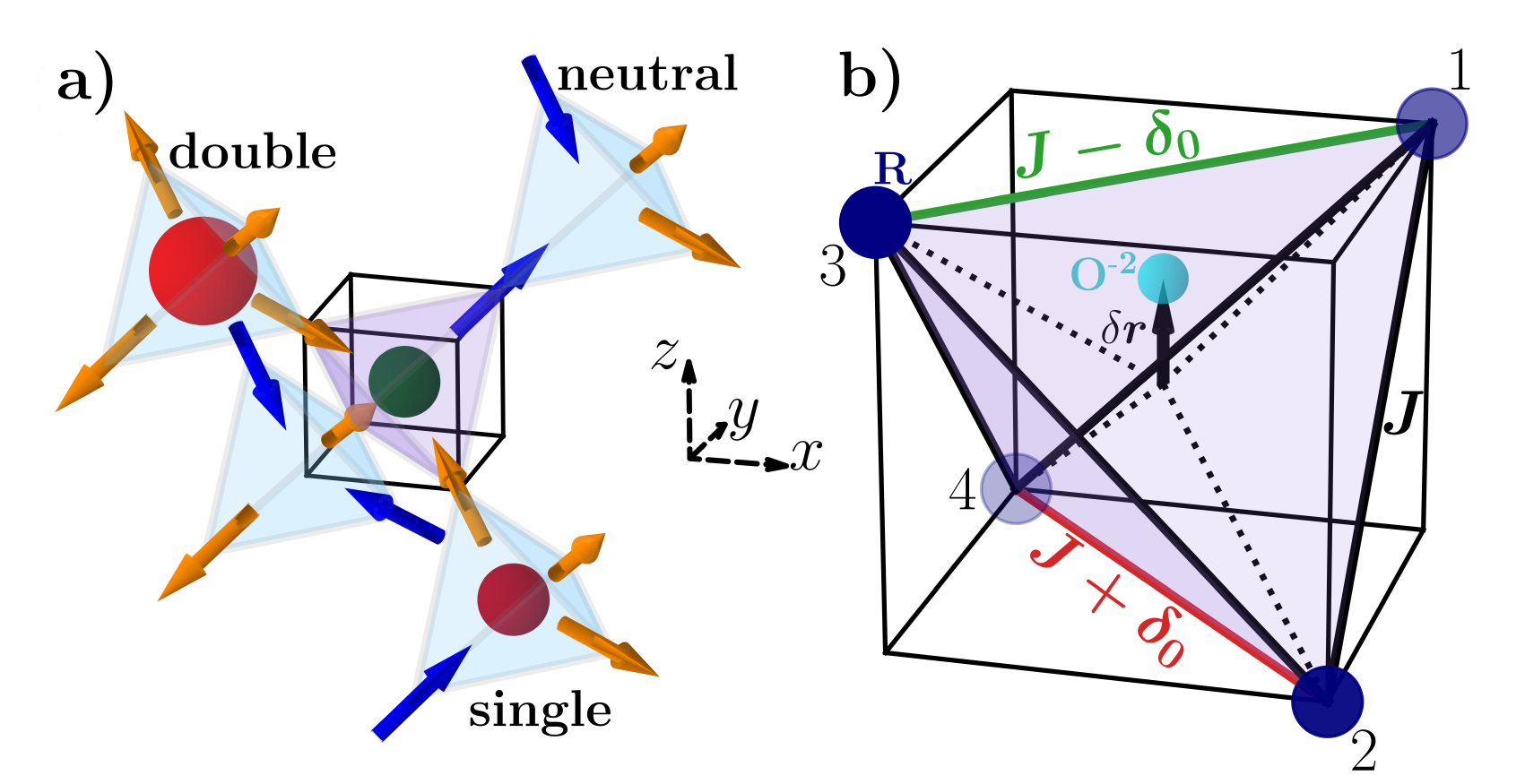}
    \caption{\textbf{Structure, magnetic monopoles and \OO-distortions.} \textbf{a)} Pyrochlore structure, with Ising spins in the shared vertices of up (embedded in a cube) and down (light blue) tetrahedra. The three-in/one-out configuration in the up tetrahedron has associated a positive single monopole in its centre (small green sphere); we also show a negative single and double monopole (small and big red spheres, respectively) and two neutral (two spins in/two-out) sites.  \textbf{b)} The displacement $\delta \textbf{r}$ of the central oxygen ion (cyan sphere) toward the $+z-$link decreases the corresponding exchange constant value (green line); the opposite exchange constant ($-z-$link, red line) is increased by this distortion, and the other four (black lines) $J$ values remain unchanged. }
    \label{figTetra}
\end{figure}

We will study a pyrochlore oxide lattice with Ising spins of the type \R. Spins will generally be associated with rare earth ions (R), while M is typically a transition metal (Ti, Sn, Zr)~\cite{Onoda2011exchange,Tomasello2018correlated,Sazonov2013} but could also be Ge or Si~\cite{zhou2011high}. The spins sit in the corners $i=1...4$ of up-tetrahedra (coloured purple in Fig.~\ref{figTetra}a)). They point along the $\langle$111$\rangle$ directions, either towards (with pseudospin variable $S_{i}=1$) or against ($S_{i}=-1$) the centre of the tetrahedron they belong to. In order to better describe the variety of states of matter we are going to study, it will be useful to employ the language of magnetic excitations or ``monopoles". We will use these two terms to refer to the topological charge even in the absence of long range dipolar interactions~\cite{Castelnovo2008}. One can group the different spin configurations of a single tetrahedron into sets, using the net entrant spin flux as a label that defines the magnetic charge~\cite{Castelnovo2008} of that tetrahedron. The same definition is valid for up or down tetrahedra. A crucial observation is that fixing the magnetic charge in a tetrahedron does not necessarily  define the spins variables in a unique way. There are six different ``neutral" or ``spin ice" configurations, with two spins pointing in and two pointing out (empty tetrahedra in Fig.~\ref{figTetra}a)). There are four positive (negative) single monopoles of charge $Q$ ($-Q$), with three spins pointing in and one out (three out and one in); they are drawn as small green (red) spheres in Fig.~\ref{figTetra}a). Finally, for double monopoles each charge identifies a single configuration: $2Q$ ($-2Q$) when all the spins point in (out) of the tetrahedron; a negative double monopole is represented by a big red sphere in Fig.~\ref{figTetra}a).

Each tetrahedron in the pyrochlore lattice can be embedded in a cube. The six links between nearest neighbour spins lie diagonally along the six faces of the cube and can be labelled using the perpendicular Cartesian axes (e.g. $+z$ and $-z$ for the links between spins $S_1$--$S_3$ and $S_2$--$S_4$, shown in green and red respectively in Fig.~\ref{figTetra}b)). Following other studies~\cite{Onoda2011exchange,Tomasello2018correlated,Sazonov2013,Jaubert2015prb} we assume that the superexchange between R-ions takes place through the oxygen ion \OO, sitting at the centre of the tetrahedra (see Fig.~\ref{figTetra}b)). In order to simplify our model for magnetoelastic coupling we will only consider the independent displacement of these non-magnetic ions, keeping all the rest at fixed positions. The restoring force for the oxygen points towards the centre of the tetrahedron and is taken to be isotropic and proportional to the oxygen's relative displacement $\delta \textbf{r}$ (see Fig.~\ref{figTetra}).   With these considerations, and taking into account only nearest neighbor magnetic interactions, our model Hamiltonian can be written as
\begin{equation}\label{H1}
{\cal H} = \sum_{\{\tetrahedron\}} \Big(\frac{1}{2} K \delta \textbf{u}^2 + \frac{1}{2} \sum_{i \neq j = 1}^4 J^{ij}(\delta \textbf{u})~ S_{i}S_{j} \Big). 
\end{equation}
Here $\delta \textbf{u} \equiv \delta \textbf{r}/r_{nn}$, with $r_{nn}$ the nearest neighbor distance, $K$ is the elastic constant for the oxygen ions. The sum runs over all (up and down) tetrahedra. $J^{ij}(\delta \textbf{u})$ is the displacement-dependent nearest neighbors superexchange energy associated to each pair. It can also be labeled using the link name $J^{\pm m}(\delta \textbf{u})$, with $m = x,y,z$ (for example, $ J^{+z}\equiv J^{13}$ for the up tetrahedron in Fig.~\ref{figTetra}b); for more details on the notation see Supplementary Information I).

For small deviations $\delta \textbf{u}$, the superexchange constants can be expanded around the undistorted value~\cite{pili2019two}, $J_0$, which corresponds to the configuration where the O occupies the central position and is thus identical for all directions. We will assume that the main effect of the O displacement over the exchange constants comes through the change in the bond angle of the R-O-R  ~\cite{Khomskii2012,Jaubert2015prb}. The net result of the angular distortion on $J$ is to make it more antiferromagnetic or ferromagnetic according to the Goodenough-Kanamori-Anderson rules.  As shown in Supplementary Information I, to first order $J^{\pm m}(\delta \textbf{u})$ is only affected by the $m$-component of $\delta \textbf{u}$: $J^{\pm \eta m}(\delta \textbf{u}) = J_0 - (\pm) \eta \tilde{\alpha} \delta u^m$, where $\eta$ takes the value $+1$ ($-1$) for up (down) tetrahedra. The constant $\tilde{\alpha} \equiv \left| \frac{\partial J^{\pm m}}{\partial \delta u^m} \right|_{\delta \textbf{u}=0}$ is the coupling constant of the global system that correlates the lattice and magnetic degrees of freedom. 

Within this approximation it is possible to recast the Hamiltonian (see Supplementary Information I) into a compact vectorial form, where the dependence on the \OO\ lattice distortion is explicit. We call this extension of the simplest spin ice model the Magnetoelastic Spin Ice model (MeSI); for zero magnetic field it is given by:

\begin{eqnarray}\label{HH}
{\cal H} \approx {\cal H}_{{\tiny \rm MeSI}} \equiv \sum_{\{\tetrahedron\}} \Big( 
&\frac{1}{2}& 3 J_{ml}^{-1}\; \delta \tilde{\textbf{u}}^2 
- \eta\ \delta \tilde{\textbf{u}} \cdot \textbf{[}S,\tilde{S}\textbf{]} _-\nonumber \\
&+& {\bf J}_{0} \cdot \textbf{[}S, \tilde{S}\textbf{]}_+
\Big).
\end{eqnarray}

Here we have defined $J_{ml} \equiv 3 \tilde{\alpha}^2 / K$ and $\delta\tilde{\textbf{u}} \equiv \tilde{\alpha} \delta \textbf{u}$, both measured in Kelvin; the sum runs along the diamond lattice, $ {\bf J}_{0} = (J_0,J_0,J_0)$ and the vectors $\textbf{[}S, \tilde{S}\textbf{]}_\pm$ have components 

\begin{eqnarray}
[S,\tilde{S} ] _\pm^x &\equiv& S_1S_{2} \pm S_{3}S_{4}  \nonumber \\ 
\small[ S,\tilde{S} \small] _\pm^y &\equiv& S_{1}S_{4} \pm S_{2}S_{3} \\
\small[S,\tilde{S}\small]_\pm^z &\equiv& S_{1}S_{3} \pm S_{2}S_{4}\nonumber.
\end{eqnarray}

The first term of Eq. \ref{HH} is the elastic energy, and it is easy to see that the last one is the usual nearest-neighbour Hamiltonian with isotropic exchange constants.  If different types of nearest neighbour bonds were to be considered (as we will do when considering the effect of magnetostriction in Section~\ref{sec:DobLay}) the latter would be replaced by a sum involving the exchange constant matrix $J^{ij}_0$: $ \sum_{i \neq j} J^{ij}_0 S_{i}S_{j}$. 

The middle term in the MeSI model is central to this work, as it contains the (linearized) magnetoelastic coupling. While the coupling constant is somewhat hidden inside $\delta\tilde{\textbf{u}} = \tilde{\alpha} \delta \textbf{u}$, we will soon show that the new energy scale $J_{ml}$ (proportional to the square of the coupling constant $\tilde{\alpha}$) is a convenient measure of the relative stability of single monopoles, the atomic-like building blocks of the new exotic phases we will study in the following sections. Also, and equally  important, it indicates how strongly magnetism will be reflected in structural properties and measurements.

\subsection{Stabilisation of a dense fluid of single monopoles: the Monopole Liquid.} \label{sec:ML}

Models of interacting entities, even simple ones, are seldomly exactly solvable. It is then a surprise that the three-dimensional MeSI model turns out to be analytically solvable for $J_0 = 0$.  Completing squares in $\delta \tilde{\textbf{u}}$, the Hamiltonian can be decomposed into an ``elastic" and a ``magnetic" term.  The first one is 

\begin{equation}\label{Helas}
{\cal H}_{\rm el} = \sum_{\{\tetrahedron\}} \frac{1}{2} 3 J^{-1}_{ml} (\delta {\textbf{O}})^2 + {\rm const.},
\end{equation}

where the components $\delta O^m =\delta\tilde{u}^{m}-\frac{J_{ml}}{3}[S,\tilde{S}]_-^m$ can be interpreted as the relative displacement of the oxygen with respect to its equilibrium position along the different axes.  Due to the magnetoelastic coupling this position depends now on the specific spin configuration in each tetrahedron. This term is quadratic and can be easily integrated out. If we include a Zeeman term, proportional to the magnetic field $\textbf{h}$ (measured in Kelvin), the effective magnetic term under a magnetic field then becomes

\begin{equation}\label{Heff}
\begin{aligned}[b]
    {\cal H}_{\rm eff}(\{S\},\textbf{h}) = \sum_{\{\tetrahedron\}} \Big( J_{ml} \prod_{i=1}^4 S_i + \frac{1}{2} J_0 \sum_{i \neq j = 1}^4 S_{i}S_{j} + \\ + \textbf{h} \cdot \sum_{i = 1}^4 \textbf{S}_{i}   \Big).
\end{aligned}
\end{equation}

The last two terms are the nearest neighbour spin-ice Hamiltonian under an applied magnetic field (with uniform exchange constant $J_0$) . For a strong magnetoelastic coupling, the first term (with a four-spin product) stabilizes a Monopole Liquid at low temperatures~\cite{Slobinsky2018}. It is easy to check that the range of stability is given by $J_0 < J_{ml} $ for positive $J_0$ (which otherwise corresponds to a spin ice phase), and $J_{ml} > -3J_0$ for negative $J_0$ (usually leading to a double monopole crystal).  

The Monopole Liquid for $J_0=0$ has been shown to be
a perfect paramagnet, with no spin correlations at any temperature\cite{Slobinsky2018}. Its ground state holds a massive residual entropy, and is equally populated by the 8 possible monopole configurations. The four-spin model (i.e., ${\cal H}_{\rm eff}$ for $\textbf{h}=0$ and $J_0=0$) was solved exactly by Barry and Wu ten years before the discovery of Spin Ice~\cite{Barry1989}. In recent years it had been suggested the possibility that lattice distortions could stabilize dense monopole phases~\cite{Sazonov2012_double,Jaubert2015spin,Slobinsky2018}; the MeSI model crystallises this idea in a clean and straightforward fashion, with the added benefit of an analytical solution.

\begin{figure}
    \includegraphics[width=0.95\columnwidth]{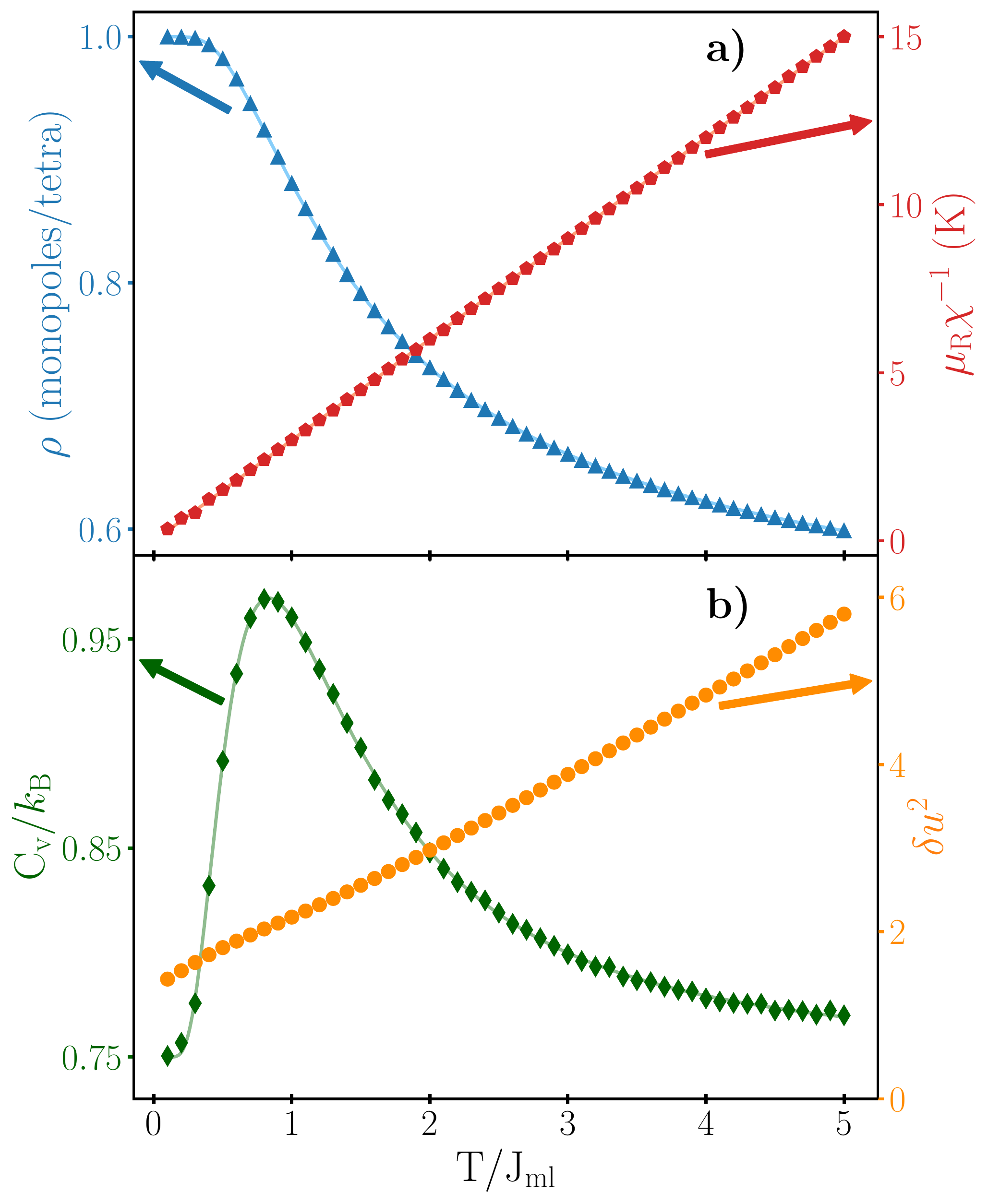}
    \caption{\textbf{Monopole Liquid.} MC simulations for the full MeSI model of Eq.~\ref{HH} (symbols), and analytical results from Barry and Wu for the effective model of Eq.~\ref{Heff} for $J_0=0$ and $\textbf{h}=0$ (lines). The cubic lattice size we simulated had $N$=8192 spins and $N/2$ moving oxygens; $J_{ml}=3\tilde{\alpha}^2/K=1K$, to fix the energy scale. \textbf{a)} As noted in Ref.~\cite{Slobinsky2018}, the spin part behaves as a perfect paramagnet; there are no spin correlations at any temperature, with a perfect Curie-law for the inverse susceptibility $\chi^{-1}$. The density of monopoles $\rho$ saturates at 1 single monopole per tetrahedron for $T<<J_{ml}$ (i.e., forms a ``Monopole Liquid"). \textbf{b)} The specific heat per unit spin $C_V$ and the mean quadratic deviation $\delta u^2$ evidence that we are really dealing with a composed system. Due to the elastic contribution ${\cal H}_{el}$ from the oxygen ions, we needed to add a constant term $3/4$ to the analytical solution to match the simulations. $\delta u^2(T)$ is not just a straight line, owing to the dependence of the O equilibrium position with the spin configuration in $\delta \textbf{O}$.}
    \label{figMC}
\end{figure}

Fig.~\ref{figMC} shows results of our Monte Carlo simulations (symbols) for the full MeSI Hamiltonian for $J_0 = 0$ (Eq.~\ref{HH}). They are compared with the exact results obtained by Barry and Wu~\cite{Barry1989}, displayed as full lines. Note that, unlike the case of  Ref.~\citenum{Slobinsky2018}, the model here involves both the spins and the (coupled) elastic degrees of freedom.  We define the density of monopoles, $\rho$, as the average number of single monopoles per tetrahedron (note that it does not count double charges). Fig.~\ref{figMC} shows that $\rho(T)$ saturates at $\rho =1$ monopole per tetrahedron for $T/J_{ml}<<1$, as expected for a dense phase of single charges; on the other hand, the  inverse magnetic susceptibility $\chi^{-1}$ is that of a paramagnet, with no evidence of increasing magnetic correlations with decreasing $T$ (Fig.~\ref{figMC}a)).  In both cases there is very good agreement between the simulations of the full model and the analytical results for the effective model~\cite{Barry1989}. On the other hand, the specific heat per unit spin $C_V$ and mean square deviation $\delta u^2$ (Fig.~\ref{figMC}b)) make apparent  that we are in fact dealing with a composite magneto-elastic system.
The solution of Barry and Wu for $C_V/k_B$ needs an extra constant term of $3/4$ to take into account the elastic energy of the $N/2$ oxygen ions, as expected from the equipartition theorem. Although according to this same theorem one would naively expect a straight line for $\delta u^2$ vs $T$, there is a kink for $\delta u^2$ in Fig.~\ref{figMC}b) for $T$ below the maximum in $C_V$.  It is a sign of coupling between the degrees of freedom: as we mentioned, the oxygen equilibrium position depends on the local spin configurations (Eq.~\ref{Helas}).

The interplay between the nearest-neighbours spin ice term proportional to $J_0$ and the four-spins term favouring a monopole liquid has been explored in Ref.~\citenum{Slobinsky2018}, where the four spin term in Eq.~\ref{Heff} was proposed as a model Hamiltonian. In addition to the Zeeman term included in Eq.~\ref{Heff} (studied in the next sections) it is interesting to consider interactions between monopoles (as those that would arise by including magnetic dipolar interactions between spins~\cite{Castelnovo2008}). A simple way to introduce nearest neighbour repulsive or even attractive forces between like-monopoles consists in including second and third nearest neighbours spin interactions with a carefully chosen ratio~\cite{Ishizuka2013shellyfish,Udagawa2016}. In order to preserve generality, we will express the interaction directly in terms of the monopolar charge on a tetrahedron, $Q_{\tetrahedron}$:

\begin{equation}\label{qq}    
    {\cal H}_{\rm QQ} = \sum_{\langle \tetrahedron,\tetrahedron' \rangle} \gamma Q_{\tetrahedron}Q_{\tetrahedron'},
\end{equation}

\noindent where $Q_{\tetrahedron}=0,\pm1,\pm2$. Here the sum runs over nearest neighbour tetrahedra, and $\gamma$ measures the strength of like-charge repulsion ($\gamma>0$) or attraction ($\gamma<0$).

We have referred to the liquid phase with a single monopole per tetrahedron ($\rho=1$) and no spin correlations as ``the'' Monopole Liquid, ML. However, other monopole liquids can be obtained by applying an external magnetic field~\cite{Slobinsky2019polarized}, changing the monopole composition, or the spin or charge correlations~\cite{Slobinsky2018}.  Some of these phases show particular patterns in the structure factor that have led to different monikers.   ``Half moons" or ``split rings" have been observed in the structure factor at the ``jellyfish point"~\cite{Udagawa2016} or the ``spin slush" phase~\cite{Rau2016}, with single monopole density $\rho \approx 0.35$ and attraction between like-monopoles. We have calculated the neutron structure factor $I^{Spin}(\boldsymbol{k})$ (see Supplementary Information III for details) for the ML in the presence of nearest neighbour attraction between like charges as per  Eq.~\ref{qq} with $\gamma < 0$ ($\rho=1$, $T<<|\gamma|<J_{ml}$). The diffuse pattern we obtain can be understood as the merging of the different half moons observed in Refs.~\citenum{Udagawa2016} and~\citenum{Rau2016}, as their features widen due to the higher density. While half moons are usually detected at finite energy~\cite{yan2018halfmoons,guitteny2013}, the ML with like-attraction is a new instance (together with Refs.~\citenum{Udagawa2016,Rau2016,mizoguchi2017PRL,mizoguchi2018PRB}) of a ground state with this feature. Interestingly, the need for an attraction between like monopoles will arise again when studying \Tb~(Section~\ref{sec:DobLay}), a compound which also shows half moons in its neutron structure factor~\cite{guitteny2013,Fennell2014magnetoelastic}.

With their exotic excitations~\cite{Castelnovo2008,morris2009}, topological phase transitions~\cite{moessner03,fennell2007pinchpKagome,Jaubert2008Kaste,Jaubert2009,Jaubert2010Multicriticality,baez16}, peculiar dynamics~\cite{Jaubert2009nat,snyder2001spin,snyder2004low,mostame2014tunable,Slobinsky2010,paulsen2014quenches}, power law correlations leading to pinch points~\cite{bram2001pinch,Fennell2009coulombphase,fennell2007pinchpKagome,Henley2010,sen2013,twengstrom2020screening}, the possibility to tune new ordered or disordered phases using magnetic fields~\cite{sakakibara2003,moessner03,hiroi2003ferromagnetic,higashinaka2003,melko2004,higashinaka2005field,sato2006ferromagnetic,sato2007field,castelnovo2012annual,grigera2015intermediate,Borzi2016}, spin ices have showed a wealth of interesting physics. In the same way, the opportunity to stabilize a completely different phase with massive residual entropy in an Ising pyrochlore opens the door to new and non-trivial forms of dense monopole matter. Part of these phases have been theoretically speculated on~\cite{Castelnovo2008,Borzi2013,Bbartlett2014,Jaubert2015spin,Jaubert2015prb,Udagawa2016,Slobinsky2018,Guruciaga2016,Raban2019multiple,Slobinsky2019polarized} or inferred through experiments~\cite{Sazonov2012_double,Borzi2016,Lefrancois2017}.  In what follows, we analyze how to obtain from the MeSI model some of these states, which, even within the realm of classical systems, do not exhaust all the possibilities opened by the inclusion of the magnetoelastic coupling.

\subsection{The Fragmented Coulomb Spin Liquid (FCSL): correlated magnetic and dipolar electric fluctuations.} \label{sec:MonCrys}

There exist previous experimental realisations and theoretical proposals for the single Monopole Crystal with the Zincblende structure (ZnMC, for short), stabilised at low temperatures by means of  extrinsic~\cite{sakakibara2003,krey2012first,Guruciaga2016,Guruciaga2019,Slobinsky2019polarized} or internal fields~\cite{Bbartlett2014,Jaubert2015spin,Lefrancois2017,cathelin2020fragmented}, or dynamical constraints~\cite{Borzi2013}.
Within this context, it was first established that spins in a crystal of single monopoles at zero field could still fluctuate~\cite{Borzi2013}. Brooks-Bartlett and collaborators noted that these partially ordered spins fragmented into two independent parts~\cite{Bbartlett2014}. A static divergence-full part was related to the monople crystal, and the remaining  (divergence-less) fragment, with neutron pinch points~\cite{Bbartlett2014,Jaubert2015spin}, characterised a Coulomb phase~\cite{Henley2010}. 
A number of Fragmented Coulomb Spin Liquids (FCSL) have been recently achieved experimentally in a pyrochlore lattice~\cite{Lefrancois2017,cathelin2020fragmented}. There, the Ir sublattice orders antiferromagnetically, acting as an effective field (with staggered values on up and down tetrahedra) over the spins in the other pyrochlore sublattice  (Ho and Dy, respectively).

Returning to our work, the inclusion of an effective monopole attraction between $+$ and $-$ charges ($\gamma > 0 $ in Eq. \ref{qq}), implicit, for example, on dipolar spin interactions, will transform the fluid of single monopoles studied in Section~\ref{sec:ML} into a ZnMC on decreasing temperature. As studied before~\cite{Bbartlett2014}, this phase would show magnetic moment fragmentation, with pinch points in the diffuse structure factor. However, in contrast with previous 
cases, there should now be spontaneous symmetry breaking between the two sites of the diamond lattice. The staggered charge density $\rho_S$ (defined as the modulus of the total magnetic charge due to single monopoles in up tetrahedra per sublattice site per unit charge) is the order parameter of the transition, which has a complex phase diagram~\cite{Borzi2013,dickman1999phase}.

\begin{figure}[htp]
    \includegraphics[width=0.5\textwidth]{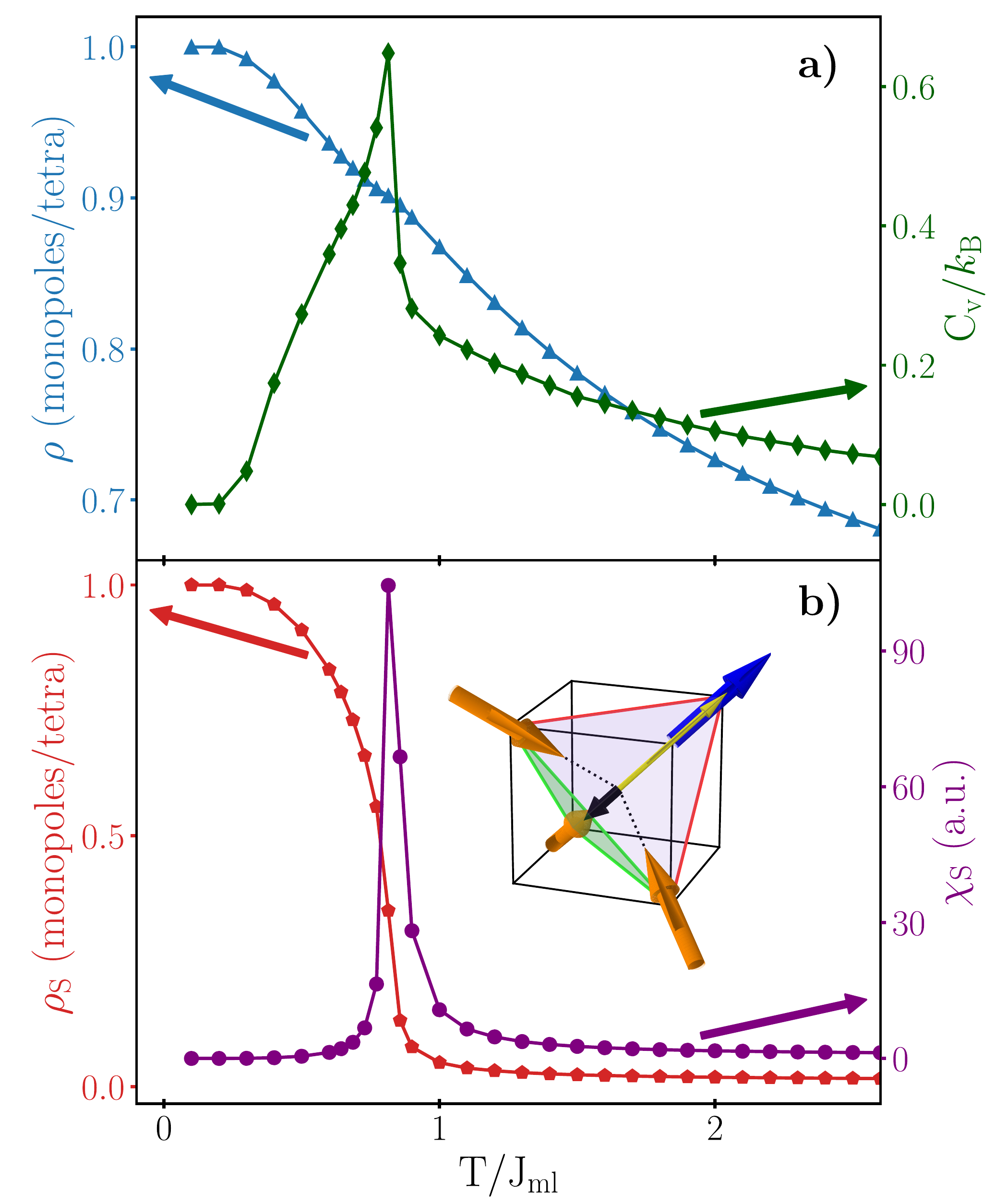} 
    \caption{\textbf{Zincblende Monopole Crystal ($J_0 = 0$ and $\gamma/J_{ml}=0.2$).} \textbf{a)} Specific heat ($C_V$) and density of single monopoles ($\rho$) along the crystallisation transition. \textbf{b)} The order parameter for crystal formation (the staggered monopole density, $\rho_S$) and its fluctuations $\chi_S$ in the same temperature range. We have subtracted the contribution of the pure vibrational degrees of freedom from $C_V$. The peak in $\chi_S$ reflects the spontaneous symmetry breaking. The inset to panel b) shows a magnetoelastic configuration of minimum energy for a positive single monopole in an up tetrahedron. The total magnetic moment (thin yellow arrow) points along the minority spin (blue arrow); it is one among a total of 8 different directions: 4 associated with positive and 4 with negative single monopoles, or 1 per spin configuration. The short black arrow corresponds to $\delta \textbf{r}$, the O$^{-2}$ displacement of minimum energy for this magnetic configuration. There are only 4 different displacements $\delta\textbf{r}$, since they do not invert under time reversal, and they are always directed towards the face of the tetrahedron with 3 antiferromagnetic (i.e. in-out) links, coloured green here. } 
    \label{fig_crystal}
\end{figure}

Fig.~\ref{fig_crystal} shows Monte Carlo simulations for the MeSI model with $J_0 = 0$ and opposite sign attraction ($\gamma/J_{ml}=0.2$) in Eq.~\ref{qq}.  We observe a high density of monopoles in the whole temperature range, saturating at $\rho=1$ at low $T$. The peak in the specific heat reflects the formation of a crystal (panel a)). Contrary to previous studies, ~\cite{Lefrancois2017,Raban2019multiple}, the abrupt increase in $\rho_S$, together with the peak in its fluctuations shows that this time the symmetry between up and down tetrahedra is spontaneously broken at the transition (Fig.~\ref{fig_crystal} panel b)). By varying the value of $J_0 > 0$ the whole phase diagram $\rho$ vs. $T$ for charges in a lattice obtained for electric charges in a lattice~\cite{dickman1999}, and then for conserved magnetic monopoles ~\cite{Borzi2013} can now be understood as emerging from a classical Hamiltonian with physical foundations. Including a negative $J_0$, a double monopole crystal can also be stabilized~\cite{denHer2000,Guruciaga2014}. Defining the total density of monopoles $\rho_T$ to include double monopoles ($0\leq \rho_T \leq 2$), a complex phase diagram would then be obtained comprising three different ground states:  the vacuum of monopoles with $\rho_T=0$, the crystal of single monopoles for $\rho_T=1$ (both exponentially degenerate and with an associated gauge field, if no other interactions are added), and the zero-entropy crystal of double monopoles for $\rho_T=2$.

Even if we take it as a possible route to the relatively unexplored physics of ``condensed monopole matter", we would not be making justice to the MeSI model if we do not consider in more detail the new, structural degrees of freedom.  As we will see below, this allows us to make apparent some of the consequences of fragmentation from a different perspective.
As sketched in the inset to Fig.~\ref{fig_crystal}b), when monopoles are stabilized at low temperatures the oxygen ions tend to be displaced along the $\langle$111$\rangle$ directions, towards one of the four triangular faces of the tetrahedron. This is the face that contains the three antiferromagnetic-like links (out-out, or in-in), painted green in Figs.~\ref{fig_crystal}b) and ~\ref{figSazo}b). It is easy to check that the O$^{-2}$ displacements $\delta\textbf{u}_i$ of a positive single monopole points antiparallel (parallel) to the total magnetic moment $\bm{\mu}_i$ of an up (down) tetrahedron. On reversing time the magnetic charge and dipole moment invert their direction, but the displacement $\delta\textbf{u}_i$ (and the electric dipole moment) remains fixed. For a given monopole charge, then, electric and magnetic moments flip in unison.

We can then argue that since the divergence-less part of the magnetic moment fluctuates like a gauge field, with neutron scattering pinch points in its structure factor~\cite{Bbartlett2014,Lefrancois2017}, the same should be true for the dipolar electric moment sitting in each tetrahedron. If this were true, aside from the usual Bragg peaks associated with the pyrochlore crystal, pinch points related to correlated oxygen fluctuations around the centre of the tetrahedron could in principle be detected as diffuse scattering using simply an electron beam, or x-ray diffraction. 
The chances to observe the effect depends critically on the magnitude of the \OO\ displacement. We will discuss this in Section~\ref{sec:Discussion}, where we also summarize the structure factor results for this and two other phases.

The fact that electronic dipolar magnetic moments could give birth to magnetic charges, and that these magnetic entities have associated electric dipolar moments has been mentioned as a further remarkable example of symmetry between electric and magnetic charges~\cite{Khomskii2012}. Another layer  of complexity is thus added by noting that the correlated fluctuations of these dipolar electric and magnetic moments lead to twin gauge fields, that could be measured by probes coupling either to electric charge or to magnetic moments.

\subsection{Double-layered crystal of single monopoles} \label{sec:DobLay} 

\begin{figure}[htp]
    \includegraphics[width=1.00\columnwidth]{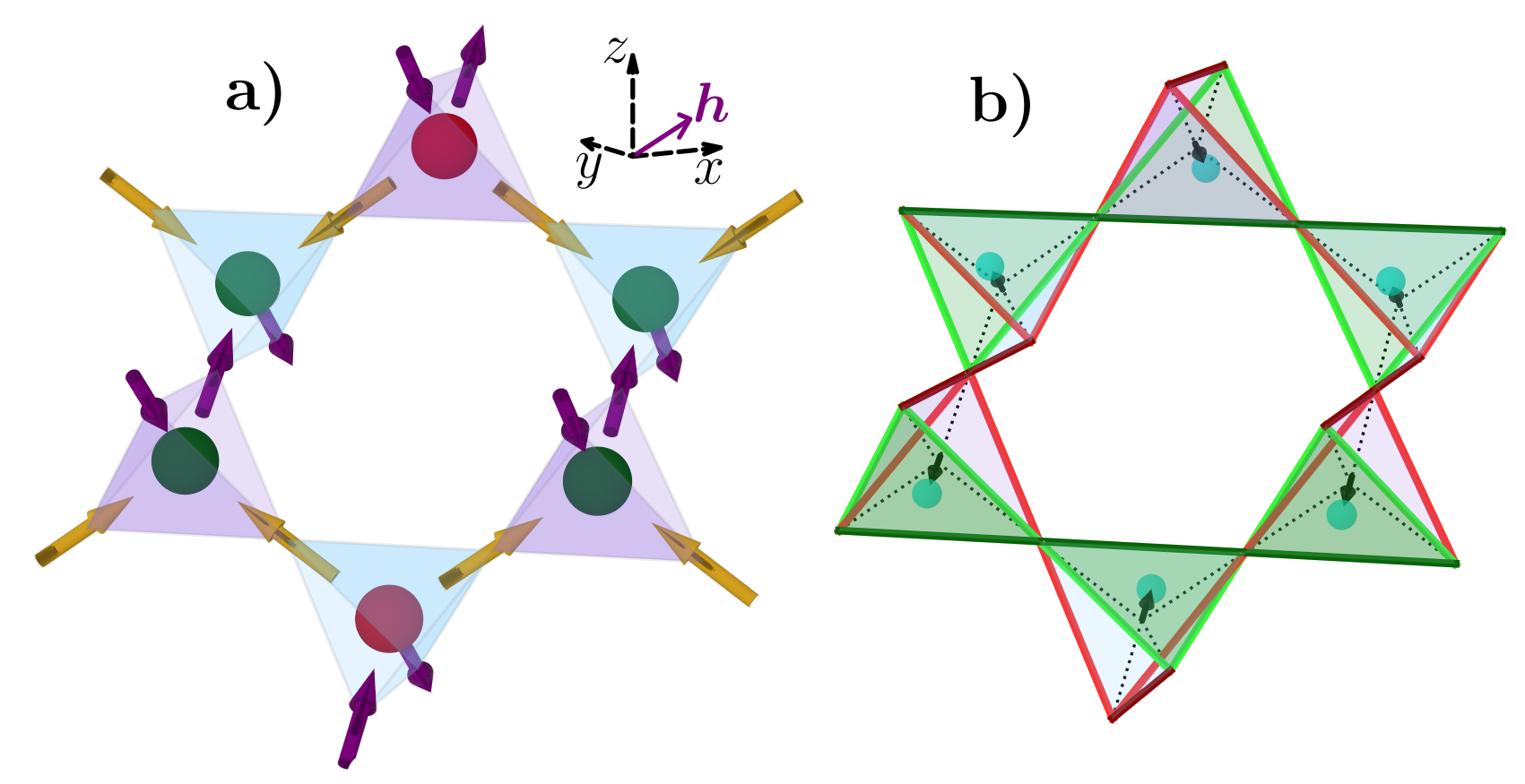}
    \caption{\textbf{Double-layered crystal of single monopoles.} \textbf{a)} Spins (drawn as arrows) and the resulting single monopole configuration (green and red spheres, indicating positive and negative charge). The spin lattice is divided in: \textit{i-} $\alpha-$spins (painted purple) polarized by $\textbf{h}||$[110], and \textit{ii-} $\beta-$spins (yellow) perpendicular to $\textbf{h}$ and building up antiferromagnetic $\beta-$chains (in-in/out-out). Monopoles linked by $\beta$-spins alternate in sign, forming $\beta_Q-$chains of charge. $\alpha-$spins decouple consecutive $\beta_Q-$chains, but an effective nearest-neighbors attraction between like monopoles (Eq.~\ref{qq} with $\gamma < 0$) can stabilize order such that the charge in two tetrahedra linked by an $\alpha-$spin is the same. \textbf{b)} Central oxygen configuration. Monopoles are stabilized by the displacement of the central oxygen (small cyan spheres) along $\langle$111$\rangle$ directions. These distortions  (which should be compared with those proposed on Fig. 6 in Ref.~\citenum{Sazonov2012_double}) decrease the exchange energy along the three bonds of the triangular face (painted green) approached by the O ions, favouring ``three-in", or ``three-out" configurations in these triangles (see a)). The spins form ``one-in/one-out" configurations on the other bonds (painted red), where the exchange constant increases. The exchange constants along $\pm z$ (painted in slightly darker colors) are further modified by magnetostriction, which triggers the O displacement.}
    \label{figSazo}
\end{figure}

Among the complex physics of \Tb~there is a clear experimental fact: upon applying an external field $\textbf{h}$ parallel to the [110] crystallographic direction, an order of alternate double layers of positive and negative monopoles is induced perpendicular to [001] \cite{Sazonov2012_double,Ruff2010} (see Fig.~\ref{figSazo}a)). To justify this charge order, Jaubert and Moessner~\cite{Jaubert2015prb} explored a classical model. The mechanism involved the long range interactions between the electric dipoles associated with single magnetic monopoles \cite{Khomskii2012}, and in a much lesser degree magnetic dipolar interactions. They found a transition from the antiferromagnetic ``all-in/all-out" phase into the bi-layer when applying a [110] magnetic field.  

Our MeSI model constitutes an alternative to this first intrinsic mechanism ever proposed to stabilize a single monopole phase~\cite{Jaubert2015prb}. While it can obviously not take into account all the complexity observed in \Tb~\cite{Taniguchi2013variabilityTb,Molavian2007dynamically,Fennell2012,Sazonov2013,Yin2013,Fritsch2014}, it is an improvement over this previous proposal, since now both magnetic and elastic degrees of freedom are considered in the same footing. Furthermore, the model provides a unified explanation for the ground state observed at zero field (a Coulomb phase~\cite{Fennell2012,Petit2012}), the double layered monopole crystal measured at moderate fields~\cite{Sazonov2012_double} (correcting the previously proposed \OO\ lattice distortions), and suggests a connection with the presence of ``half moons" in neutrons diffuse scattering at finite energy~\cite{guitteny2013,Fennell2014magnetoelastic}.

The application of a strong magnetic field along [110] does not fully order the  Ising pyrochlore lattice. Spins on $\alpha$-chains ---running along [110], represented by purple arrows in Fig.~\ref{figSazo}a)--- are completely polarized at low temperature. This results in effectively decoupled spins in $\beta$-chains (yellow arrows in the figure), with magnetic moments perpendicular to $\textbf{h}$. Only four possible spin configurations are then possible in each tetrahedron. Two of these are spin ice-like, with no average O displacement~\cite{hiroi2003ferromagnetic}; the other two are a positive and a negative single monopole, leading to the antiferromagnetic $\beta-$chains of spins, and $\beta_Q-$chains of alternating charge shown in Fig.~\ref{figSazo}a)~\cite{Guruciaga2016,Guruciaga2019}. Unlike the figure (chosen to show a double monopole layer ordering) these $\beta-$chains are not coupled by the spin structure and --unless an explicit energetic coupling is included-- would lead to an incoherent arrangement of $\beta_Q-$chains.

Before tackling the question of charge coherence, the non-trivial question we should answer concerns magnetic charge stability. Why would \Tb\ change its ground state under a magnetic field  $\textbf{h}||$[110] from a subset of the 2-in 2-out manifold to that of a dense phase of single monopoles~\cite{Sazonov2012_double} when the field is strong enough? 
Since the component of the magnetic moments along $\textbf{h}$ is the same for the two chosen single monopoles and for the neutral sites, the response cannot rely on the Zeeman energy alone. It is interesting to note firstly that if we impose the alternate oxygen displacements along $z$-axis proposed by Sazonov et al (Fig. 6 in Ref.~\citenum{Sazonov2012_double}), the MeSI model naturally leads to a dense phase of non-coherent $\beta_Q-$chains of magnetic monopoles. The only requisite is a displacement big enough to overcome the energy associated to the usual nearest neighbors term proportional to $J_0$. Alternatively we will now inquire on the effect of $\textbf{h}$ on the lattice structure, and then of that on the spin lattice through magnetoelastic coupling $\tilde{\alpha}$.

We are not the first to notice the possible importance of the giant magnetostriction observed for $\textbf{h}||$[110]~\cite{Ruff2010magnetoelastics} for stabilizing the double-layered monopole phase in \Tb~\cite{Jaubert2015prb}. Here we will include its effect implicitly in the MeSI model through the exchange matrix $J^{ij}_0$ in Eq.~\ref{HH}. Adding a Zeeman term for $\textbf{h}||{\rm [110]}$ the extended MeSI model can be written as:

\begin{equation}\label{HH110}
\begin{aligned}[b]
    {\cal H}_{\tiny{\rm{MeSI}}}^{{\tiny \rm [110]}} = \sum_{\tetrahedron} \Big( \frac{1}{2} 3 J_{ml}^{-1} (\delta \tilde{\textbf{u}})^2 -\eta\ \delta \tilde{\textbf{u}} \cdot \textbf{[}S,\tilde{S}\textbf{]}_-  + \\ +  \frac{1}{2} \sum_{i \neq j = 1}^4  J^{ij}_0 S_{i}S_{j} + \textbf{h} \cdot \sum_{i=1,3} \textbf{S}_{i} \Big).
\end{aligned}
\end{equation}

Rare-earth ions usually have associated a very strong spin-orbit coupling; through it, the torque acting on spins can affect the orbital angular momentum, and then the lattice. Based on symmetry~\cite{Cao2009field110} the effect of the field along [110] on the exchange constants is modelled through $J^{13}_0=J_0^{+\eta z}=J_0 - \delta(h)$ and $J^{24}_0=J_0^{-\eta z}=J_0 + \delta(h)$; for simplicity, we keep the other exchange constants $J^{ij}_0$ unchanged (see Fig.~\ref{figTetra}b)). In order to detect the formation of a dense phase of monopoles in our Monte Carlo simulations we measure $\rho$ and two more specific quantities: 
$i-$ the average of the staggered O displacement along the $z-$axis, $\langle \delta u^z_{\rm stagg}\rangle$, that is sensitive to the O-displacement proposed by Sazonov and collaborators. We compute it as the average of $\delta u^z$ on up tetrahedra minus that on down tetrahedra (see Fig.~\ref{figSazo}a)); and $ii-$ the order parameter, $OP$, for the double-layer crystal of single monopoles, calculated as the staggered charge on [100] planes made of up-tetrahedra. If we call $Q^{up}_j$ the total charge in the $j-$th [100] plane of up tetrahedra, the $OP$ is computed as:

\begin{equation}\label{OP}
 \begin{aligned}[b]
    OP =  \Bigg|\sum_{j=1}^{2L} (-1)^j Q^{up}_j \Bigg|,
 \end{aligned}
\end{equation}

where $L$ is the linear size of the system and we are counting two planes of up tetrahedra per unit cell.

\begin{figure}[htp]
	\includegraphics[width=0.95\columnwidth]{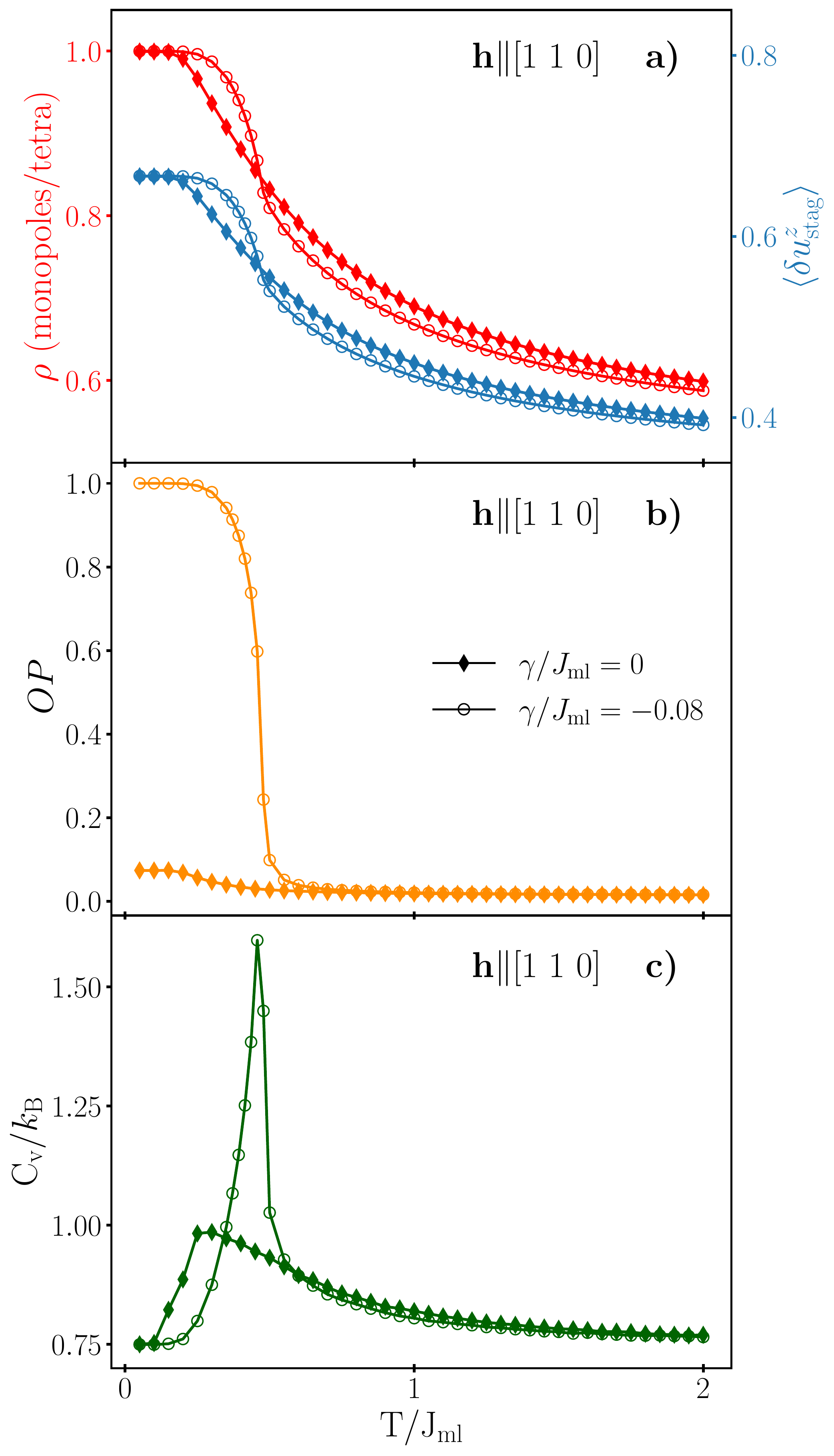}
	\caption{\textbf{Double-layered crystal of single monopoles.} Monte Carlo simulations for the extended MeSI model (Eq.~\ref{HH110}) as a function of temperature and fixed field $\textbf{h}||$[100], with $h/J_{ml}= 13.4$ and $L=8$. In order to guarantee a spin ice phase at zero field, $J_0/J_{ml}= 1.1$, $\delta(0)/J_{ml}=0$, and $\delta(h)/J_{ml}=-0.5$. Curves with filled symbols where obtained using Eq.~\ref{HH110}, while those with open symbols are the equivalent after adding an effective charge-charge interaction (Eq. \ref{qq}) with $\gamma/J_{ml}=-0.08$. \textbf{a)} Monopole density $\rho$ (left) and average staggered O displacement along $z-$axis, $\langle \delta u^z_{\rm stagg} \rangle$ (right). \textbf{b)} Order parameter. \textbf{c)} Specific heat. The inclusion of some attraction between like charges (Eq.~\ref{qq}) leads to an spontaneaus symetry breaking transition leading to the magnetic and structural phase shown in Fig.~\ref{figSazo}}.
	\label{figMC_DoubleL}
\end{figure}

Fig.~\ref{figMC_DoubleL} shows the results obtained for the complete MeSI model of Eq.~\ref{HH110} as a function of temperature (filled symbols). We used a fixed field $h/J_{ml} = 13.4$, with $\delta(h)/J_{ml}=-0.5$. In order to guarantee a spin ice phase at zero field we set $J_0/J_{ml}= 1.1 > 1$. The condition to destabilize the spin ice state in favour of a monopole phase at zero temperature is $\delta(h) < J_{ml} - J_0$. With $J_0/J_{ml}=1.1$, we make sure that a two-in--two-out state is favoured for $h=0$ ($\delta(0)=0$), compatible with the observed Coulomb phase in \Tb. On the other hand, the value $\delta(h)/J_{ml}=-0.5$ ensures a single monopole phase at a finite field. We see the density of single monopoles saturating smoothly at $\rho=1$ below $T/J_{ml}=0.2$. Since the intensity of the magnetic field was chosen in order that the $\alpha-$spins would be saturated (and thus the magnetisation) for $T/J_{ml} < 1.4$ this increase in $\rho$ involves only the (antiferromagnetic) arrangements of $\beta-$spins. We can see that the staggered average of the O displacement along $z-$axis, $\langle \delta u^z_{\rm stagg}\rangle$, follows closely this behaviour, showing that the O-displacement along $z-$axis seem to coincide with that predicted in ref. ~\citenum{Sazonov2012_double}. The negative value we needed to use for $\delta(h)$ is quite encouraging: it means that $J^{ij}=J^{+\eta z}$ in the link parallel to the field increases with field, while the one perpendicular ($J^{-\eta z}$, painted green in Fig.~\ref{figSazo}b)) decreases. The crystal contracts along the [110] field and expands in the direction perpendicular to it, in full accordance with the observed distortions under magnetic field~\cite{aleksandrov1985,Ruff2010magnetoelastics}.

In spite of the above, we notice that the specific heat $C_V$ (Fig.~\ref{figMC_DoubleL}c), full symbols) shows only a broad Schottky anomaly on decreasing temperature, while the order parameter $OP$ varies very little. This tells us that the spin ice-like ground state has changed into a dense monopole phase of incoherent $\beta_Q-$chains, producing no spontaneous symmetry breaking. It is easy  now to see that an effective interaction like the one in Eq.~\ref{qq} with attraction between like charges ($\gamma < 0 $) is the coupling needed to obtain the double monopole layer structure, since it favours the positive (negative) charges in contiguous $\beta_Q-$chains to be next to each other (Fig.~\ref{figSazo}a)). It can be the result of second and third nearest neighbours exchange interactions~\cite{Ishizuka2013shellyfish}, and may be related to the ``half moons" in \Tb\ neutron scattering experiments~\cite{guitteny2013,Fennell2014magnetoelastic}.
Alternatively, the additional term can also be thought as an effective way to include the effect of the electric dipolar interactions, that have been proved to lead to the double-layered monopole crystal~\cite{Jaubert2015prb}. It is important to stress that their role here is not to stabilize single monopoles~\cite{Jaubert2015prb}, but (more subtly) to favour a particular monopole arrangement.

The open symbol curves in Fig.~\ref{figMC_DoubleL} show the marked changes we measured after adding the monopole interaction term (Eq.~\ref{qq}) with $\gamma/J_{ml}=-0.08$ to the extended MeSI model. We can see that $\rho$ and $\langle \delta u^z_{\rm stagg}\rangle$ reach saturation in a much sharper way. The abrupt jump in $OP$ (reaching the value of 1), and the peak in $C_v$ (with an extra area under it) show that these sharp features are connected with the spontaneous symmetry breaking by the double monopole layer structure.  In addition to the spin and monopole configurations, displayed on Fig.~\ref{figSazo}a), our model provides the lattice distortions linked to this magnetic structure (Fig.~\ref{figSazo}b)). As discussed before (Figs.~\ref{fig_crystal}b) and~\ref{figSazo}b))
and differently from Ref.~\citenum{Sazonov2012_double}, this displacement is not only vertical: the O ion tends to approach the triangular surface of each tetrahedron where the three spins point likewise (darkened in the figure), so as to reduce the value of the exchange constants along the corresponding links. 

Given the big magnetic moments associated with Tb$^{+3}$, a brief consideration is needed regarding dipolar magnetic interactions. As discussed in Ref.~\citenum{Jaubert2015spin}, their effect will be twofold. Firstly, the preference of these interactions for two-in/two out states should be compensated by the huge magnetostriction of \Tb\ (i.e., the transition into a dense monopole phase would occur at higher fields/deformations than if no magnetic dipolar forces were included). Secondly, dipolar interactions would disfavour the proximity of like magnetic charges, demanding bigger values of $|\gamma|$ (i.e., bigger next-nearest neighbours exchange interactions, or dipolar electric moments).

\section{Discussion}\label{sec:Discussion}

It is interesting to compare the resulting structures for some of the ground states which combine a maximum density of single monopoles and extensive residual entropy. We can now put together three pieces of information: the usual (spin) magnetic scattering, scattering from the (distorted) \OO-lattice, and hypothetical scattering from magnetic charges. They were calculated from simulations at very low temperature, so that magnetic excitations are negligible and \OO-ions are displaced only along the unit cell diagonals (see Section~\ref{sec:MonCrys}, and Supplementary Information II and III for details on the simulations and the precise definitions used in the Structure Factor calculations).

 \begin{figure*}
      \centering
        {\includegraphics[width=\textwidth]{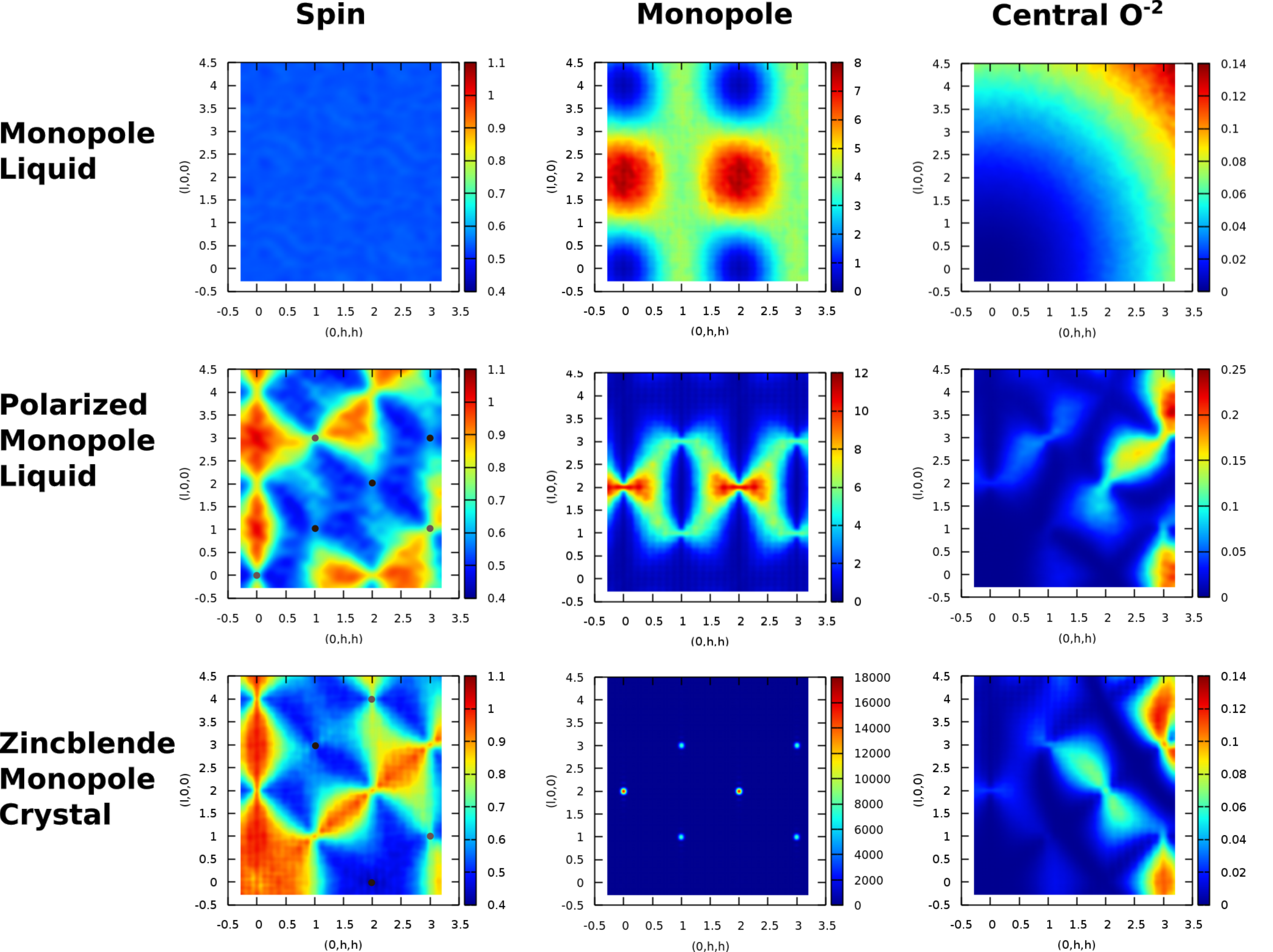}} 
      \caption{\textbf{Structure factors for the different disordered phases.} The scattering centre (Spins, Single Monopoles, or the oxygen ions near the centres of the tetrahedra) varies along the horizontal axis (see Supplementary Information III for details on the structure factor calculation). The names of three of the studied phases are indicated on the left, with increasing charge order (while not necessarily spin order) progressing downwards. We only show diffuse scattering for O$^{-2}$ ions. Bragg peaks from the fragmented divergence-full component of the magnetic moment in the Zincblende Monopole Crystal (ZnMC), or by the spins aligned by the applied magnetic field along [100] in the Polarized Monopole Liquid (PML), are indicated schematically by full circles. Three different gauge field types can be observed, associated to spins, magnetic charges, and oxygen displacements. While the Monopole liquid does not show pinch points in any case, the PML is most remarkable, reflecting the existence of a Coulomb phase in the three channels. The Fragmented Coulomb Spin Liquid~\cite{Bbartlett2014,Lefrancois2017,cathelin2020fragmented} should reflect the existence of a Coulomb phase also through lattice distortions.
      }
    \label{fig_resumen}
 \end{figure*}

Fig.~\ref{fig_resumen} shows a comparison of the calculated structure factors within the [$h,l,l$]  plane  of reciprocal space. Three of the monopole phases (including the Polarizad Monopole Liquid --PML-- studied in detail in reference~\citenum{Slobinsky2019polarized}) run along the rows of the table. The ``scattering centres'' (spins, magnetic charges, and the O$^{-2}$ ions displaced from the centre of each tetrahedron) run along its columns. For the O$^{-2}$ displacement we show only the diffuse part, removing the trivial contribution from the regular diamond lattice formed by the \OO\ average position and $k$-dependent charge (see Supplementary Information III). On the other hand, Bragg scattering peaks due to static spins (polarised by the field or associated to the curl-free part of the magnetic moment in the crystal of single monopoles) are indicated schematically by full circles.

Monopole order progresses downwards in this table, as illustrated by the second column: broad maxima for the Monopole Liquid give place to pinch points in the PML and then to sharp Bragg peaks for the Zincblende Monopole Crystal. Regarding the Monopole Liquid, in spite of the maxima in the monopole channel, it shows no spin-spin correlations at all, which is also true for the \OO~displacements (first row in Fig.~\ref{fig_resumen}). The existence of these peaks in the charge channel may be counterintuitive given the total spin decorrelation. Monopole-monopole correlations in the ML are due to construction constraints, due to the underlying spins~\cite{Guruciaga2014,Slobinsky2018}.

As previously mentioned in the text, the Zincblende Monopole Crystal shows pinch points both in the spin and the \OO~channel; strong Bragg peaks reflect the monopole correlations in the crystal. As the ML, the Polarized Monopole Liquid (middle row) has no spin or monopole long range order
~\cite{Slobinsky2019polarized}. Notably, and differently from its unpolarized version, the PML has associated a gauge field that can be related either to spins, magnetic charges or displaced \OO-ions. In principle, radiation interacting with any of these three particles could show the pinch points characterizing a Coulomb phase. 
The ability to detect effects related to the electric dipole on monopoles depends on its magnitude. While there are indications of the presence of such electric dipoles in \Tb\ and \Dy~\cite{grams2014,jin2020experimental}, the O displacement $\delta r$ has not been measured. Within our model, we can obtain it through $J_{ml}$ (Eq.~\ref{Helas}), provided we know the coupling constant $\tilde{\alpha}$ and the elastic constant $K$. A rough estimation based on the experimental and numerical results obtained in Refs. \cite{edberg2019dipolar,edberg2020effects,gupta2009lattice,kushwaha2017vibrational} gives a small $J_{ml}$ for \Dy\ and \Ho, on the order of $10$ mK. In turn, this leads to $\delta r \approx 0.1$ pm for these canonical spin ices.
On the other hand, Jaubert and Moessner~\cite{Jaubert2015prb} estimate a bigger $\delta r$ for \Tb (within the picometer range), similar to that observed in multiferroic materials. While this is still considerably small, new methods based on traditional XRD have been very recently proposed and used to observe distortions within this range in a strontium titanate oxyde \cite{richter2018picometer}. The chances of a direct observation of the magnetoelastic phenomena we propose can increase if the efforts are first concentrated on compounds with a big coupling between magnetic and lattice degrees of freedom, starting with monopole crystals. The double monopole layer in \Tb\ could then be an excellent place to begin.

In summary, we have introduced an extension to the usual Hamiltonians used for studing Ising spin systems on pyrochlore oxides \R. The Magnetoelastic Spin Ice (MeSI)  model  includes the spin coupling to the lattice of central \OO-ions in up and down tetrahedra, through the dependence of the superexchange constant $J(\delta\textbf{u})$ on the oxygen displacement ($\delta\textbf{u}$). We showed that, in the strong coupling limit, lattice distortions turn single monopoles (the excitations of the spin ice materials) into actual building blocks for novel ground states with maximum density of magnetic charges.
Crucially, $\delta\textbf{u}$ works as a dynamic, internal field; there is thus no explicit symmetry breaking, and all eight single monopoles are \textit{a priori} equally probable in each tetrahedron. 

This avenue to new ground states and novel physics is widened by an additional factor: the \OO~distortion implies an electric dipolar moment. This means that the distortions $\delta \textbf{u}$ are not just ``hidden" degrees of freedom that allow for the occurrence of new phases, but can be thought as probes to investigate the underlying magnetism, or, in a multiferroic fashion, to control the material.

We have presented some examples of the above. The first one is a Monopole Liquid ground state, stabilized for the first time with a Hamiltonian which physical bases. Including an attraction between magnetic monopoles of the same charge leads to ``half moons" features in the spin structure factor of this liquid; this makes a direct link to the ``spin slush" phases~\cite{Udagawa2016,Rau2016}. Remarkably, notwithstanding its simplicity, the MeSI model provides a unified framework that explains the zero field ground state measured in \Tb~\cite{Fennell2012} and the double-layered monopole crystal at moderate fields~\cite{Sazonov2012_double}; this includes an improved version for the distortion on the \OO-lattice proposed previously~\cite{Sazonov2012_double}.
The classical treatment of distortions at low temperatures, albeit  unrealistic, can be understood as a simple way to probe the magnetoelastic instabilities of this system.
At zero field, the MeSI model recreates the phase diagram for single monopole spontaneous crystallization studied in~\cite{Borzi2013,Bbartlett2014} without resourcing to artificial constraints, and would allow to extend it to include double monopoles~\cite{Guruciaga2014}. The spontaneous crystallization of a dense liquid of single monopoles into the Zincblende structure gave rise to the Fragmented Coulomb Spin Liquid~\cite{Bbartlett2014,Jaubert2015spin,Lefrancois2017}. As stressed before, there is a close parallelism between some electric and magnetic phenomena in frustrated Ising pyrochlores~\cite{Khomskii2012}. Our access to the elastic degrees of freedom provides another layer of complexity to this, by showing that the \OO~displacement $\delta\textbf{u}$ in the FCSL phase is related (like the spin) to a gauge field. 
Perhaps more singular is the case of the Polarised Monopole Liquid~\cite{Slobinsky2019polarized} (i.e., the monopole liquid with an applied magnetic field along [100]). This disordered state is a Coulomb phase from the point of view of three different degrees of freedom:  spins, magnetic monopoles and elastic distortions. As with the FCSL, pinch points could be detected using diffuse neutron scattering or simply x-ray or electron diffraction.

Although here we have concentrated mainly on the magnetic degrees of freedom and on the strong coupling limit, the MeSI model opens perspectives of research in other grounds. For instance, the weak coupling regime can be used to describe in a combined way (spins and lattice distortions) some of the physics of spin ice materials~\cite{saito2005magnetodielectric,katsufuji2004,Khomskii2012,grams2014}: their true ground state \cite{melko2001,pomaranski2013,Borzi2016,bhattacharjee2016acoustic,samarakoon2020machine,henelius2016,edberg2019dipolar}, the effect of uniaxial pressure~\cite{mito2007,edberg2019dipolar}, and the new phases that appear under applied field \cite{sato2007field,grigera2015intermediate,Borzi2016,Lin2013halfmag,baez16}.

\section{Acknowledgements}
This work was supported by Agencia Nacional de Promoci\'on Cient\'\i fica y Tecnol\'ogica (ANPCyT) through grants PICT 2013-2004, PICT 2014-2618 and PICT 2017-2347, and Consejo Nacional de Investigaciones Cient\'\i ficas y T\'ecnicas (CONICET) through grant PIP 0446. Part of this project was carried out within the framework of a Max-Planck independent research group on strongly correlated systems.

\section*{Supplementary Information}

\subsection{Derivation of the Hamiltonian}\label{App:recast}

Most of the results shown in this work have been derived from Eq. 2 of the main text, obtained after rewriting the magnetoelastic exchange energy for small \OO distortions $\delta \textbf{u}$ under certain assumptions. Here we will make explicit this procedure. The energy we are concerned with corresponds to the second term in Eq. 1 of the main text; for a single tetrahedron it can be written as $\frac{1}{2} \sum_{i \neq j = 1}^4 J^{ij}(\delta \textbf{u})~ S_{i}S_{j}$.

\begin{figure}[htp]
    \centering
    \includegraphics[width=0.6\columnwidth]{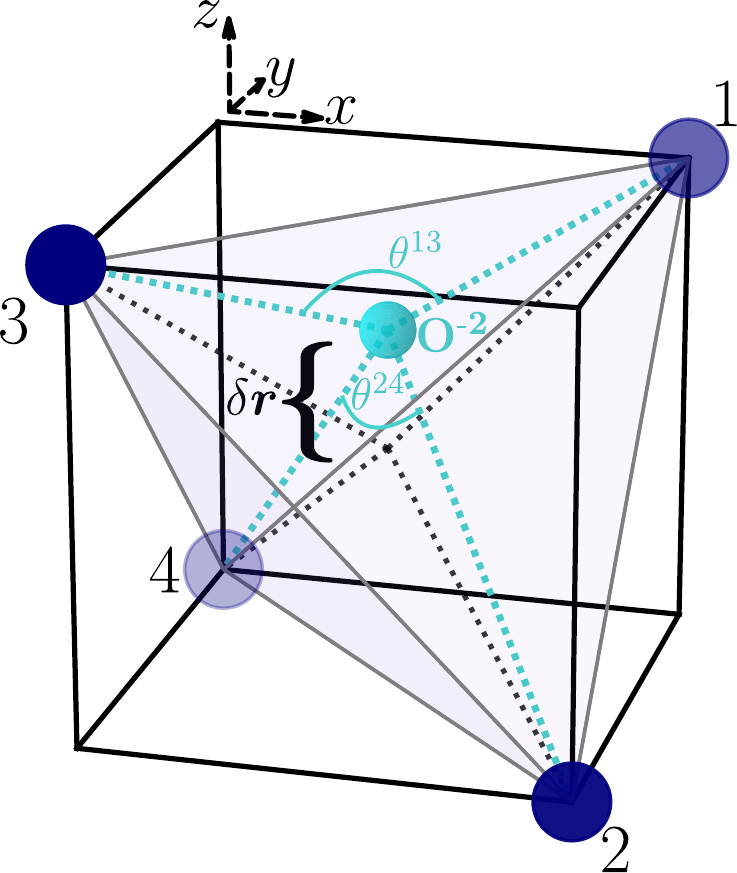}
    \caption{\textbf{Relation between \OO~distortions and exchange angles for an up tetrahedron ($\eta=1$).} For an undistorted \OO\ position the four angles are identical. When, for instance, the central oxygen moves along $z$, the upper angle $\theta^{13}$ becomes more obtuse, making the superexchange energy $J^{+z}=J^{13}$ more negative (i.e., more antiferromagnetic in terms of whole spins). The opposite happens with $\theta^{24}$ and $J^{-z}=J^{24}$.} 
    \label{figApendix1}    
\end{figure}

Note that the $\{S_i\}$ are pseudospins; a positive value of the superexchange constant $J^{ij}$, for instance, would favour pairs of pseudospins of different sign, which translates into a ferromagnetic-like arrangement in the tetrahedron (one spin in, the other pointing out)~\cite{bramwell2001sci}. As justified in the text, we assume that the superexchange interaction between the nearest-neighbour spins $S_i$ and $S_j$ is mediated by the \OO-ion near the centre of the spin arrangement, and that it is a function of the angle $\theta^{ij}$ formed by the position vectors of the spins relative to this ion, $J^{ij}=J(\theta^{ij})$ (Fig.~\ref{figApendix1}). We will now obtain a functional dependence of the couplings $J^{ij}$ on the displacement of the oxygen $\delta \textbf{r}$, for very small displacements ($|\delta \textbf{u}| \equiv |\delta \textbf{r}|/r_{nn} \ll 1$); we assume, for simplicity, that the spins remain in fixed positions of the lattice.  

We concentrate first on an \textit{up} tetrahedron.  When the oxygen is at the central position, the $\theta^{ij}$ angles are all equal to $\theta=\arccos(-\frac{1}{3}) =109.47 \degree$. To lowest order, the cosine is linear in $\delta\theta$ when expanded around this value. 
Let us, for instance consider the angle $\theta^{13}$ between the relative position vectors of $S_1$ and $S_3$ with respect to the displaced oxygen (see Fig.~\ref{figApendix1}). The positions involved (measured from the exact centre of the tetrahedron, in units of $r_{nn}$) are $\textbf{u}_1=\sqrt{2}/4\times(1,1,1)$ for spin 1; $\textbf{u}_3=\sqrt{2}/4\times(-1,-1,1)$ for spin 3, and $\delta \textbf{u}=(\delta u_x, \delta u_y, \delta u_z)$ for the \OO-ion. The expression for $\theta^{13}$ can be obtained via the scalar product between its defining vectors: $\textbf{u}_1-\delta \textbf{u}=(\sqrt{2}/4-\delta u_x, \sqrt{2}/4-\delta u_y, \sqrt{2}/4-\delta u_z)$, and $\textbf{u}_3-\delta \textbf{u}=(-\sqrt{2}/4-\delta u_x, -\sqrt{2}/4-\delta u_y, \sqrt{2}/4-\delta u_z)$. From this we get:

\begin{widetext}
\begin{equation*}\label{cos}
 \begin{aligned}[b]
     \cos \theta^{13}     = 
    \frac{-1/8-\sqrt{2}/2\times\delta u_z+O(\delta u^2)}
         {\sqrt{
                [(\sqrt{2}/4-\delta u_x)^2+(\sqrt{2}/4-\delta u_y)^2+(\sqrt{2}/4-\delta u_z)^2 ]
                [(-\sqrt{2}/4-\delta u_x)^2+(-\sqrt{2}/4-\delta u_y)^2+(\sqrt{2}/4-\delta u_z)^2]
           }
         }.
    \end{aligned}
\end{equation*}
\end{widetext}

\noindent Taylor expanding this expression to first order in $\delta u_x$, $\delta u_y$, and $\delta u_z$, it is easy to see that 
\begin{equation}\label{cosapp13}
 \cos \theta^{13}  \approx -\frac{1}{3}-C\delta u_z+O(\delta u^2), 
\end{equation} 

\noindent where $C=8^{3/2}/9$. 

In this first order approximation only the displacement along
$z$ ---which is the only coordinate that has the same value for spins 1 and 3--- remains. It may be useful to remember that if this displacement has a positive value, it yields a negative contribution to Eq.~\ref{cosapp13}, implying a more obtuse angle. Following  Goodenough-Kanamori-Anderson rules for superexchange, we can then expect that the \OO-ion displacement decreases the value of the exchange constant of the magnetic link it approaches (it makes it less ferromagnetic, or more antiferromagnetic). This is illustrated in Figs. 1b), 3 and 4b) of the main text, where we can see that pairs of spins closer to the \OO-ion (linked by green segments) tend to point all out or all in.

Proceeding analogously for the angle between spins $S_2$ and $S_4$, the common coordinate is again $z$, but this time with a negative sign; it yields a term with a positive sign:

\begin{equation*}\label{cosapp24}
 \cos \theta^{24}  \approx -\frac{1}{3}+C\delta u_z+O(\delta u^2),
\end{equation*} 

This can be repeated, for the other angles obtaining:

\begin{equation*}\label{cosapp14}
 \cos \theta^{14}  \approx -\frac{1}{3}-C\delta u_y+O(\delta u^2),
\end{equation*} 

\begin{equation*}\label{cosapp23}
 \cos \theta^{23}  \approx -\frac{1}{3}+C\delta u_y+O(\delta u^2),
\end{equation*} 

\begin{equation*}\label{cosapp12}
 \cos \theta^{12}  \approx -\frac{1}{3}-C\delta u_x+O(\delta u^2), 
\end{equation*} 

\begin{equation*}\label{cosapp34}
 \cos \theta^{34}  \approx -\frac{1}{3}+C\delta u_x+O(\delta u^2).
\end{equation*} 

\noindent For each case, only the coordinate with the same value for both spins considered produces a first order term, and this term has opposite sign with respect to  the value of the coordinate.

The same equations but with opposite sign in the second term would be obtained for a down tetrahedron (for instance,  a positive $\delta u_z$ brings the \OO~nearer the link connecting spins $S_1$ and $S_3$ in an \textit{up} tetrahedron (Fig.~\ref{figApendix1}), but it does the opposite in a \textit{down} one).

The previous calculation was purely based on geometry. Resuming physical grounds, we now expand  $J^{ij}(\cos{\theta})$ to first order in $\delta u$. We also label the exchange constant using a symbol derived from the link position respect to the Cartesian axes; for example, for an up tetrahedron $J^{+z}\equiv J^{13}$ (see Fig.~\ref{figApendix1}); on the other hand, for a down tetrahedron $+z$ indicates the link between spins $S_2$ and $S_4$. Using this, we obtain:
\begin{equation*}\label{J12}
 J^{+\eta x} \equiv J^ {12} = J_0 - \eta\; \tilde{\alpha} \delta u_x+O(\delta u^2)
\end{equation*}
\begin{equation*}\label{J34}
 J^{-\eta x} \equiv J^ {34} = J_0 + \eta\; \tilde{\alpha} \delta u_x+O(\delta u^2)
\end{equation*}
\begin{equation*}\label{J13}
 J^{+\eta y} \equiv J^ {14} = J_0 - \eta\; \tilde{\alpha} \delta u_y+O(\delta u^2)
\end{equation*}
\begin{equation*}\label{J24}
 J^{-\eta y} \equiv J^ {23} = J_0 + \eta\; \tilde{\alpha} \delta u_y+O(\delta u^2)
\end{equation*}
\begin{equation*}\label{J14}
 J^{+\eta z} \equiv J^ {13} = J_0 - \eta\; \tilde{\alpha} \delta u_z+O(\delta u^2)
\end{equation*}
\begin{equation*}\label{J23}
 J^{-\eta z} \equiv J^ {24} = J_0 + \eta\; \tilde{\alpha} \delta u_z+O(\delta u^2).
\end{equation*}

\noindent Here,  $\eta=1$ for up and $\eta=-1$ for down tetrahedra; $\tilde{\alpha}$ is the coupling constant
\begin{equation*}
\tilde{\alpha} \equiv 
\left| \frac{\partial J^{\pm m}}{\partial \delta u^m}\right|_{\delta \textbf{u}=0} \\
= C\times
\frac{\partial J^{ij}}{\partial \cos{\theta^{ij}}}\Bigr|_{\cos{\theta^{ij}=-1/3}}, \end{equation*}

\noindent assumed positive~\cite{Khomskii2012,Jaubert2015prb}. $J_0$ is the zeroth order approximation, $J_0=J^{ij}(\delta \textbf{u}=0)$, shared by all the links in the absence of structural symmetry breaking (see for instance the case of a field applied along [110], Eq.5 in Section V of the main text).

If, for instance, we sum the Hamiltonian terms for an up tetrahedron involving $\delta u_x$, we obtain $-\tilde{\alpha} \delta u_x (S_1 S_2-S_3 S_4)$, which is just the first addend of the scalar product in the second term of Eq. 2 of the main text. The addends involving $J_0$ have been gathered in the third term in this Hamiltonian, which amounts to the usual nearest neighbours spin Hamiltonian~\cite{melko2004}. The first term in Eq. 2 of the main text accounts for the elastic energy; in the absence of magnetoelastic coupling, this term ensures that the equilibrium position of the central oxygen is given by $\delta \textbf{u}= (0,0,0)$ (i.e., at the centre of the tetrahedron).

\subsection{Details on the numerical simulations.}\label{Asimul}

We provide here some details on the simulations used in the main text to study the equilibrium properties of the Ising pyrochlores. We simulated $L\times L \times L$ conventional cubic cells of the pyrochlore lattice ($16\times L\times L \times L$ spins) with Metropolis and Conserved Monopoles \cite{Borzi2013} algorithms. In all cases the boundary conditions were set to be periodic along the cubic primitive vectors. The specific heat and susceptibilities were calculated using the fluctuations of the corresponding quantities.

\subsubsection{Metropolis algorithm}
Given that the MeSI model is a composite system, we separated each step of the simulation into a single spin flip or single elastic \OO~movement. The Metropolis algorithm was then implemented, using Eq. 2 of the main text to evaluate the probability to accept or reject a given spin flip or \OO-ion movement. Typical sizes were $L=8$ ($N=8192$ spins, and $N_t=2048$ \OO-ions).
To simulate the elastic distortions we considered displacements of the \OO-ions using spherical coordinates, with $\delta u$ chosen randomly
in a distribution from 0 to a temperature dependent maximum $\delta u_{max} (T)$. The latter is introduced so as to make more efficient the algorithm~\cite{pili2019two}.  We considered the relaxation times of the magnetic degrees of freedom to be much shorter than the corresponding elastic ones, thus,  within each elastic move we performed a complete magnetic Monte Carlo step. We verified that this election has no effect on the equilibrium properties of the system. After reaching equilibrium, we averaged the data over $50$ independent runs, taking $2 \times 10^4$ time-steps at each field and temperature.
    
\subsubsection{Conserved Monopoles Algorithm}
Energy minimisation in a single monopole site requires the \OO-ion to be situated along the $\langle$111$\rangle$ diagonals, approaching the tetrahedron face with all spin pairs in antiferromagnetic-like configurations (see Fig. 3b) of the main text). This fact makes easy, in a ground state, to infer the \OO\ distortions $\{\delta \textbf{u}\}$ if the magnetic degrees of freedom $\{S_i\}$ are known. We have profited from this fact to calculate the structure factors for different dense single monopole ground states with better statistics and in larger lattices with a minimum computational effort. Using the conserved monopoles algorithm~\cite{Borzi2013,baez16} we travelled along the desired magnetic ensamble, and then deduced the elastic coordinates for each spin configuration. While each configuration had in truth two neutral sites, the density of monopoles is very near $\rho=1$ for the system sizes used (for $L=8$ the proportion between neutral sites and single monopoles is $2/4094$).

\subsection{Calculation of the different structure factors} \label{App:Scatt}

The simulated neutron structure factors have been calculated following the expression:

\begin{equation*}
    I^{Spin}(\boldsymbol{k})=\frac{1}{N}\sum_{ij}\langle S_i S_j\rangle\,\left(\boldsymbol{\mu}^{\perp}_i\cdot\boldsymbol{\mu}^{\perp}_j\right)\,e^{i\boldsymbol{k}\cdot \boldsymbol{r}_{ij}}
\end{equation*}

\noindent where $i$ and $j$ sweep the pyrochlore lattice, $N$ is the number of spins, and $\langle ... \rangle$ represents thermal average (in this case, that of the product of pseudospins at sites $i,j$). The spin quantization directions are given by $\{ \hat{\mu}_i \}$ (parallel to the $\langle 111 \rangle$ directions). Then, $\boldsymbol{\mu}^{\perp}_i$ is the component of $\hat{\mu}_i$ of the spin $\boldsymbol{S_i} = S_i \hat{\mu_i}$ at site $i$ perpendicular to the scattering wave vector $\boldsymbol{k}$: 

\begin{equation}
    \boldsymbol{\mu}^{\perp}_i=\hat{\mu}_i-\left(\hat{\mu}_i\cdot\frac{\boldsymbol{k}}{|\boldsymbol{k}|}\right)  \frac{\boldsymbol{k}}{|\boldsymbol{k}|}.
\end{equation}

As in Refs.~\citenum{Slobinsky2018} and~\citenum{Slobinsky2019polarized}, we also calculated a structure factor associated with magnetic charges, through the Fourier transform for the charge-charge correlation function:

\begin{equation}
    I^{Q}(\boldsymbol{k})=\frac{2}{N}\sum_{\alpha\beta}\langle Q_{\alpha} Q_{\beta}\rangle\,e^{i\boldsymbol{k}\cdot \boldsymbol{r}_{\alpha \beta}},
\end{equation}

\noindent where the greek indices $\alpha,\beta$ now sweep the sites of the diamond lattice where the monopoles live, $N/2$ is the number of tetrahedra, $Q_{\alpha}$ represents the topological charge at position $\boldsymbol{r}_{\alpha}$, and ${\boldsymbol{r}}_{\alpha\beta}$ is the distance between monopoles.

Finally, we have measured the diffuse structure factor associated to the displaced \OO-ions near the ground state of the system. Assuming an atomic form factor unity, we calculated: 

\begin{eqnarray}
    I^{O^{-2}}(\boldsymbol{k})=\frac{2}{N}\sum_{\alpha\beta} &&\langle (e^{i\boldsymbol{k}\cdot \delta \textbf{r}_{\alpha}}-q^{av}(\boldsymbol{k})) \times \nonumber \\ && \times (e^{-i\boldsymbol{k}\cdot \delta \textbf{r}_{\beta}}-q^{av}(\boldsymbol{k}))\rangle  e^{i\boldsymbol{k}\cdot \boldsymbol{r}_{\alpha \beta}},\nonumber
\end{eqnarray}

\noindent where $q^{av}(\boldsymbol{k})=\langle e^{i\boldsymbol{k}\cdot \delta \textbf{r}_{\alpha}} \rangle$ is an average, $k-$dependent \OO~``charge" in the perfect diamond lattice. 

These functions have been obtained after thermal averages over sets composed of $500-1000$ independent configurations for a system size $L=8$.

\bibliographystyle{apsrev4-1}
\bibliography{main}
\end{document}